\documentclass[conference, onecolumn]{IEEEtran}

\usepackage{amsmath,amssymb,amsfonts}
\usepackage{algorithmic}
\usepackage{graphicx}
\usepackage{textcomp}
\usepackage{xcolor}
\usepackage{siunitx}
\usepackage{subcaption}

\usepackage[style=ieee]{biblatex}
\AtBeginBibliography{\small}
\makeatletter
\newcommand{\linebreakand}{%
  \end{@IEEEauthorhalign}
  \hfill\mbox{}\par
  \mbox{}\hfill\begin{@IEEEauthorhalign}
}
\makeatother

\begin{document}

\title{ABO\textsubscript{3} Perovskites' Formability Prediction and Crystal Structure Classification using Machine Learning  }


\author{\IEEEauthorblockN{Minhaj Uddin Ahmad}
\IEEEauthorblockA{\textit{Electrical and Electronic Engineering} \\
\textit{Islamic University of Technology}\\
Dhaka, Bangladesh \\
minhajuddin@iut-dhaka.edu}
\and
\IEEEauthorblockN{A.Abdur Rahman Akib}
\IEEEauthorblockA{\textit{Electrical and Electronic Engineering} \\
\textit{Islamic University of Technology}\\
Dhaka, Bangladesh \\
abdurrahman22@iut-dhaka.edu}
\linebreakand
\IEEEauthorblockN{Md. Mohsin Sarker Raihan}
\IEEEauthorblockA{\textit{Biomedical Engineering} \\
\textit{Khulna University of Engineering and Technology}\\
Khulna, Bangladesh \\
msr.raihan@gmail.com}
\and
\IEEEauthorblockN{Abdullah Bin Shams}
\IEEEauthorblockA{\textit{Electrical and Computer Engineering} \\
\textit{University of Toronto}\\
Toronto, Ontario M5S 3G4, Canada \\
ab.shams@utoronto.ca \\
(Corresponding author)}
}
\maketitle

\begin{abstract}
Renewable energy sources are of great interest to combat global warming, yet promising sources like photovoltaic (PV) cells are not efficient and cheap enough to act as an alternative to traditional energy sources. Perovskite has  high potential as a PV material but engineering the right material for a specific application is often a lengthy process. In this paper, ABO\textsubscript{3} type perovskites' formability is predicted and its crystal structure is classified using machine learning with high accuracy, which provides a fast screening process. Although the study was done with solar-cell application in mind, the prediction framework is generic enough to be used for other purposes. Formability of perovskite is predicted and its crystal structure is classified with an accuracy of 98.57\% and 90.53\% respectively using Random Forest after 5-fold cross-validation. Our machine learning model may aid in the accelerated development of a desired perovskite structure by providing a quick mechanism to get insight into the material's properties in advance.
\end{abstract}

\begin{IEEEkeywords}
Photovoltaics, ABO\textsubscript{3} Perovskite, Machine Learning, Formability, Structure, Renewable Energy.
\end{IEEEkeywords}

\section{Introduction}
One of the most pressing issue of our day is global warming. Global energy demand has risen considerably, which to a large extent has been met burning fossil fuels. This creates Carbon dioxide, a major contributor to global warming. By bringing renewable energy sources into the mix, efforts are being made to reduce reliance on carbon-based conventional energy sources. Solar energy is particularly promising and researches are being conducted to find flexible and more efficient materials for PV panels that can harvest more of the sun's inexhaustible energy that falls upon the earth's surface. Solar energy is one of the building blocks of today's smart grid models. It has also found its use in ``green communication" systems~\cite{jahid2017pv, jahid2018green}. Recent works suggest that due to the advancements of technology, it is more feasible than ever to use solar energy that will provide a significant share of global energy requirements.

A perovskite is any compound with a crystal structure of ABX\textsubscript{3} where X is an anion, often halide or oxide. Most of the metallic ions can form perovskite structures. Perovskite materials are a popular research focus because of their high tolerance to defect, compositional flexibility, cation configuration, long charge diffusion lengths, high absorption coefficients, small exciton binding energies, low trap densities, etc. which allows for use in numerous applications by tuning the material’s properties. Perovskites have a higher power output whilst still being lighter in weight than traditional crystalline silicon. For solar cells, power conversion efficiency (PCE) and cost dictate their feasibility for commercialization. It has been shown that 300nm thin film is enough for absorbing all visible light above its band gap. The PCE of perovskite solar cells (PSC) reached 25.2\% in 2019 and increasing, which exceeds that of traditional thin-film solar cell~\cite{ma_realistic_2020}. Rigid solar cells are thicker and heavier than flexible PSCs, which makes them comparatively easily portable~\cite{li_recent_2018}. It can be used in small electronic gadgets including wearable electronics as well as covering large surfaces~\cite{di_giacomo_progress_2016}. Traditional crystalline silicon solar cells suffer from a complex and expensive manufacturing process that limits their large-scale application~\cite{meillaud_recent_2015}. Manufacturing of commercial PSC ($>$100cm\textsuperscript{2}) is scalable using new low temperature methods like one-step/two-step solution deposition, solvent-quenching, alternating precursor deposition, duel source vacuum co-evaporation, solid source direct contact fabrication and vapor-phase coating methods for large-scale perovskite film production~\cite{yang_recent_2019, nie_high-efficiency_2015}. Smaller ($<$1cm\textsuperscript{2}) devices can use the spin coating methods. It is also suitable with large-scale roll-to-roll fabrication, giving it an edge over conventional fabrication processes. These fabrication processes have shown success with uniform morphology, full surface coverage, pinhole-free high-quality crystal films~\cite{yang_recent_2019}. They can be easily mounted to architectures of various shapes such as curved surfaces~\cite{di_giacomo_progress_2016}. 

Outstanding structural stability shown by many of the developed ABO\textsubscript{3} materials indicates a high chance of discovering new better performant combinations. This class of perovskite oxides has a wide variation in band-gap and conductivity. The properties can be tuned by simple changes of elements in the material developing process~\cite{hao_anomalous_2014}. However, not all ABO\textsubscript{3} perovskite crystal structures of these varieties show the level of stability required for an efficient and reliable operation. As efforts are ongoing to develop more economic material, prior prediction of stability is crucial to speed up the development~\cite{fu_stability_2019}. The traditional way of developing material which depends upon trial and error, human expertise, and sometimes sheer luck; may not be the best considering the time and effort it takes. An alternative Machine Learning based approach that cruises through a sea of data to form a statistical framework for stability prediction is proven to be effective~\cite{lu_accelerated_2018}.

Machine learning makes it possible to input an enormous volume of data and present data-driven recommendations, based on advanced statistical models~\cite{lheureux_machine_2017}. The use of advanced mathematical tools like Density Functional Theory (DFT), Monte Carlo Simulation for thermodynamic stability calculation involves solving a complex quantum mechanical problem; requiring huge computational power~\cite{curtarolo_high-throughput_2013}. Machine learning (ML) provides an alternative computationally lenient approach~\cite{lu_accelerated_2018}. Moreover, even with the overhead of costlier computation, theoretical models sometimes struggle to find the material with the property required for practical operation~\cite{tao_machine_2021}.

Discovery of cost-effective material will facilitate more practical implementations, leading to a greater contribution to overall green energy production. In this regard, We opted for machine learning to classify the ABO3 perovskite structure and predict whether a composition can form perovskite or not. Moreover, Random Forest algorithm was used with 5-fold cross-validation for classifying the crystal structure. SMOTE analysis was used to balance the dataset, to make sure that the distribution of the dataset is not biased. 

\section{Methodology}
The dataset used in this work consists of 6005 ABO\textsubscript{3} perovskite-type oxides. Out of 16 features in the dataset, we have used 9 of them. The list of features that are used for training our structure classification model is given below:
\begin{tabular}{@{$\quad \bullet$ }ll}
    $r(AXII)$ & 12 site coordination of cation A \\
    $r(AVI)$ & 6 site coordination of cation A  \\
    $EN(A)$ & Electronegativity of cation A \\
    $EN(B)$ & Electronegativity of cation B \\
    $l(A-O)$ & Length of covalent A-O bond \\
    $l(B-O)$ & Length of covalent B-O bond \\
    $\Delta{ENR}$ & Difference electronegativity with radius \\
    $t_G$ & Goldschmidt tolerance factor \\ 
    $\mu$ & Octahedral factor \\
\end{tabular}

For predicting perovskite formability of a given compound, we used the following features, along with the other features mentioned above except $t_G$.
\begin{itemize}
    \item Lowest Distortion - 1 (Cubic)
    \item Lowest Distortion - 2 (Tetragonal)
    \item Lowest Distortion - 3 (Orthorhombic)
    \item Lowest Distortion - 4 (Rhombohedral)
\end{itemize}

The rest are excluded because they are literary data; explaining the crystal structure and the compound of the perovskite. Out of 6005 target values, 53 was removed from the ‘Lowest distortion’ column as they do not report any value. The ionic radii(r) describe a comparison of ionic and crystal radii with different coordination and charge states~\cite{shannon_revised_1976}. The radii of A cation for 12 site and 6 site coordination values are taken from the database~\cite{noauthor_shannon_nodate}. Other radii values of B cation for 6 coordination are also estimated from the same database. The perovskites that have been experimentally reported are selected from the literatures~\cite{Hautier_Fischer_Ehrlacher_Jain_Ceder_2011, li_formability_2004, giaquinta_structural_1994, demkov_integration_2014}. 
Perovskites undergo a local distortion from cubic structure. These distorted perovskites can have a variety of symmetries, including the rhombohedral, tetragonal, and orthorhombic distortions. The target feature for the crystal structure prediction is ‘Lowest distortion’ based on the work of Emery et al~\cite{emery_high-throughput_2017}.  

The stability and formability of different perovskite structures are heavily dependent on the Goldschmidt tolerance factor $t_G$. The tolerance factor is used as a feature that considers all ionic radii. It is well established that practically all perovskites have this value between 0.75 and 1.00. Cubic perovskites, with the exception of BaMoO\textsubscript{3}, have ‘$t_G$’ values in the range of 0.857–1.032~\cite{kumar2008prediction}. 
\begin{equation}
    t_G =  \frac{r(A)+r(B)}{\sqrt{2(r(B)+r(O))}}
\end{equation}
when the $t_G$ value is unity it indicates a perfectly cubic structure. $t_G$ greater than 1 suggests a large-sized A cation and leads to a hexagonal crystal structure formation. On the contrary, if ${t_G}<{0.8}$, it precludes the perovskite formation and other possible structures might form. 

Octahedral factor ($\mu$) is the ratio of the radius of the small cation B over the radius of anion O is vital for the stability of the perovskite. The minimum value of octahedral factor required for the formation of cubic perovskites is as low as 0.433, and the $\mu$ value of cubic perovskites is as high as 0.704~\cite{kumar2008prediction}.
\begin{equation}
    \mu = \frac{r(B)}{r(O)}
\end{equation}

Electronegativity is a measure of the tendency of an atom to attract a bonding pair of electrons. Ion's mean electronegativity ($EN$) values are taken from the data used here~\cite{pilania2016finding}. Based on those data, oxygen's ionic radius and electronegativity are assumed to be 1.40 and 3.44, respectively. The BV model is used to determine lengths of the bonds (l) of A-O and B-O. \cite{chen2017bond}.

Radius A has two features: $XII$ and $VI$ coordination. The ionic radius is dependent on the coordination number. In a perfect perovskite structure, the A-site cation's coordination number is XII. Lesser coordination numbers (VIII or VI) are frequently found in structures with poor symmetry, like orthorhombic and rhombohedral~\cite{bartel2019new}.

\subsection{Prediction Framework}
Multi-class classification is proposed for the detection of ABO\textsubscript{3} perovskite structures (cubic, tetragonal, orthorhombic, and rhombohedral). We used 5952 data points to represent the perovskite-type ABO3 crystal structure and scikit-learn python module to perform feature encoding and preprocessing. Because machine learning algorithms can only work with numerical values, categorical values are converted to numerical values. The dataset is split into training and validation sets, with 20\% of the data used to validate the trained algorithm. The dataset is split into 5 equal folds to use the cross-validation to estimate how the model will perform on unseen data. The dataset was highly imbalanced. SMOTE (Synthetic Minority Oversampling Technique) is performed on the dataset to balance it because otherwise, the minority class will have far too few examples for any model to effectively select the decision boundary. To carry out these steps, we used the scikit-learn library’s built-in functions. 

\begin{figure}[htbp]
\centering
    \begin{subfigure}{\linewidth}
    \centering
    \includegraphics[width=0.8\linewidth]{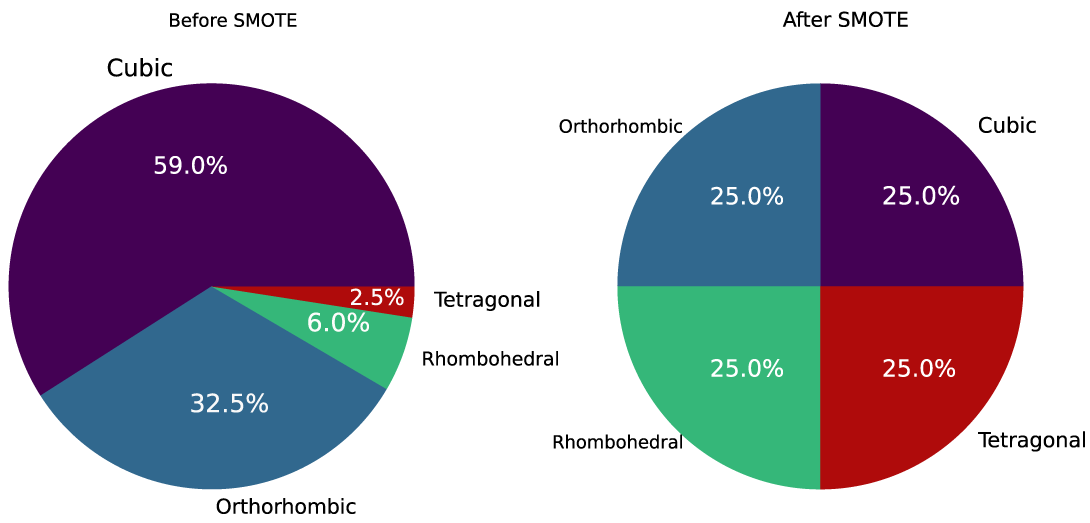}
    \caption{Before and After SMOTE for structure classification}
    \end{subfigure}
    
    \begin{subfigure}{\linewidth}
    \centering
    \includegraphics[width=0.8\linewidth]{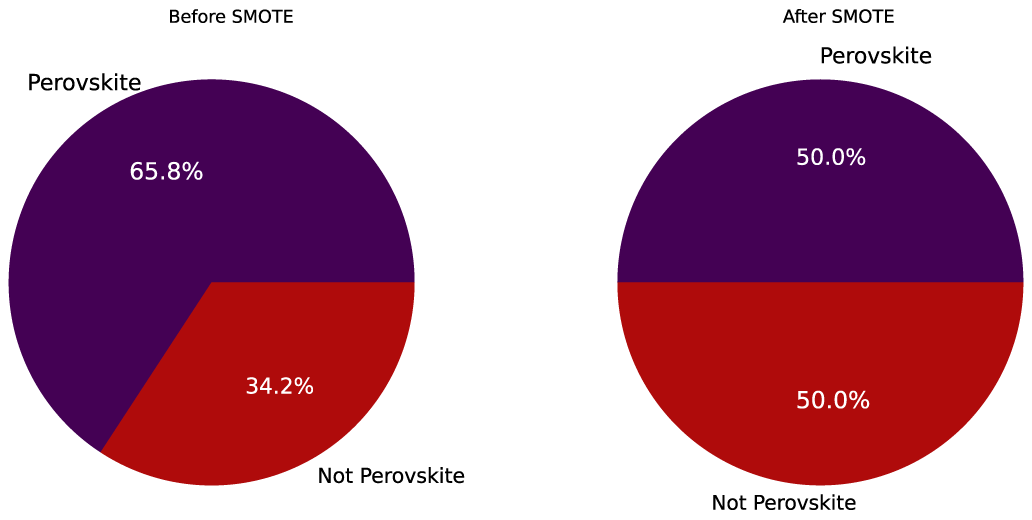}
    \caption{Before and After SMOTE for formability prediction}
    \end{subfigure}
\end{figure}

A binary classification model is proposed for the second model, where whether the crystal forms a perovskite structure or not is predicted. If a crystal structure has $t_G$ value between 0.82 to 1.10, it can form perovskites~\cite{kumar2008prediction}.  Based on the $t_G$  value, the new binary feature column titled ‘Perovskite’ is added to the dataset, and is set to be the target feature for the model. 

The second model that predicts formability of perovskite is based on the range of $t_G$ values 0.82${<}t_G{<}$1.10. We added a new binary feature column titled ‘Perovskite’ is added to the dataset using that $t_G$ range. It identifies whether the material is perovskite or not and is set to be the target feature for this model. The $t_G$ feature was dropped because it creates high feature-feature correlation as the newly added column is directly derived from $t_G$ values. We used one-hot encoding on the ``lowest distortion" feature to identify the impact of crystal structure on this perovskite formability prediction. 

\subsection{Machine Learning Algorithms}

We did a randomized search to determine a set of best-performing parameters. We implemented the randomized search using Sk-learn library's RandomizedSearchCV() class. The probability of obtaining ideal parameter is significantly higher with random search due to its pattern. This may result in the model being trained on the optimized parameters without aliasing~\cite{bergstra2012random}.

\subsubsection{Random Forest Algorithm}
Random Forest is an ensemble bagging type machine learning algorithm that falls under the category of supervised learning. The outcome of the random forest method is determined by the predictions based on decision trees. It makes predictions by taking the average of or accumulating the output of several trees simultaneously~\cite{breiman2001random}. In our implementation, the parameters chosen for perovskite structure classification are: n\_estimator=1400, criterion=`entropy' and max\_depth=40, max\_features=`auto', min\_samples\_leaf=1, min\_samples\_split=2. For perovskite formability prediction the chosen parameters are: n\_estimator=1500, min\_samples\_leaf=1, min\_samples\_split=5.

\subsubsection{XGBoost}
XGBoost is an ensemble Machine Learning technique that is based also on decision trees and employs a gradient boosting framework. It uses better software and hardware optimization to acquire better results in less computation time. ~\cite{chen2016xgboost}. In our implementation, the chosen parameters are: n\_estimators=100, gamma=0.1, learning\_rate=0.25, max\_depth=12, objective=`multi:softprob'. For perovskite formability prediction we used the default parameters.

\subsubsection{LightGBM}
LightGBM is a high-performance, distributed gradient boosting framework. This algorithm splits the tree leaf-wise, whereas other boosting methods split the tree depth- or level-wise. In LightGBM, the leaf-wise approach reduces loss more than the level-wise technique, resulting in improved accuracy. In our implementation, the chosen parameters are: learning\_rate=0.3, max\_depth=15, min\_child\_weight=3 and for perovskite formability prediction the parameters was used default parameters.

\section{Result and Discussion}
The Correlation coefficient is a statistical measure used to determine the relationship between two randomly chosen variables. It is used to determine the strength of a relationship between two random variables. Correlation coefficients values have a range of -1 to +1. A strong positive linear relationship is represented by a value close to +1, and -1 represents a robust negative linear relationship. A number near to or equal to zero indicates no correlation between the variables. Spearman’s correlation ($\rho$) is checked to ensure low correlation feature values of the dataset.

\begin{figure*}[!htb]
\centering
    \begin{subfigure}[hbt]{0.485\linewidth}
    \centering
    \includegraphics[width=\linewidth]{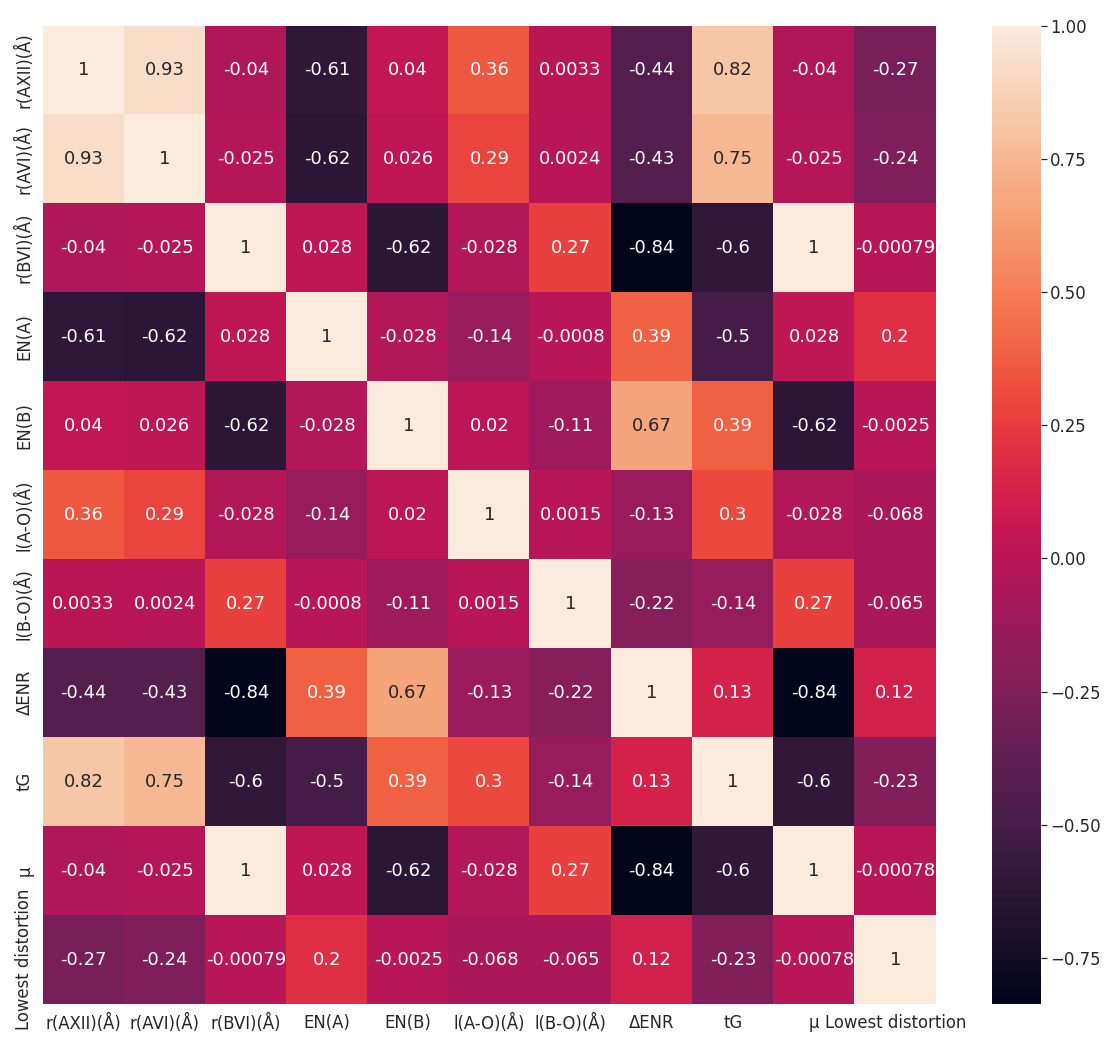}
    \caption{structure classification}
    \end{subfigure}
    \hfill
    \begin{subfigure}[hbt]{0.485\linewidth}
    \centering
    \includegraphics[width=\linewidth]{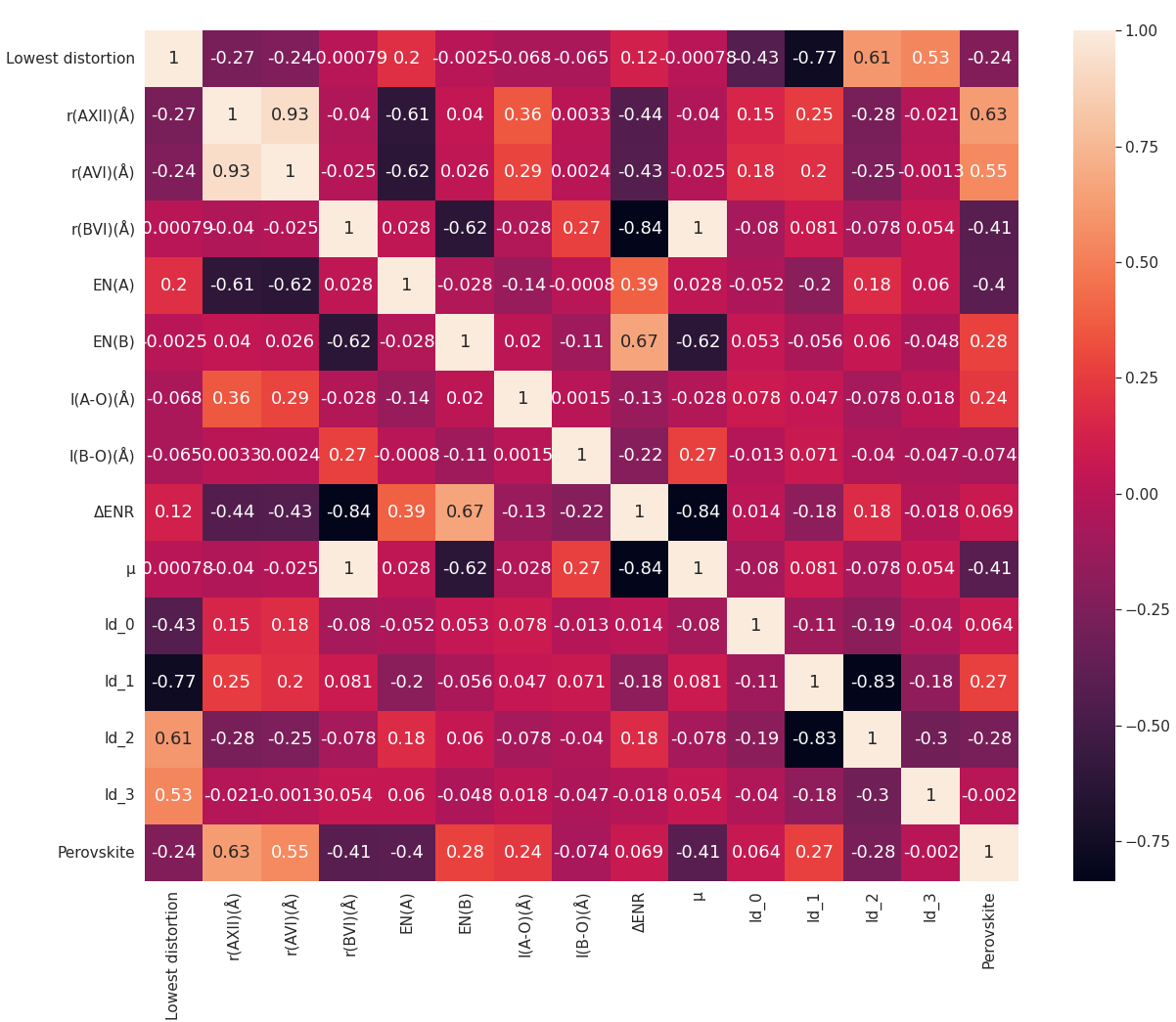}
    \caption{formability prediction}
    \end{subfigure}
\caption{Feature correlation matrices for structure classification and formability prediction}
\label{fig:shap class}
\end{figure*}

ABO\textsubscript{3} perovskite crystal structure classification and perovskite prediction both use four evaluation metrics: Accuracy, Precision, F1 score, and Recall. These are used for comparing the performance of the machine-learning algorithms~\cite{chicco2020advantages}. The evaluation metrics are computed from a confusion matrix where the values are True Positive (TP), True Negative (TN), False Positive (FP), and False Negative (FN) respectively.

It can be noted that our first classification outperformed previous studies. To make the model evaluation more robust we used k-fold cross-validation. This technique partitions the data into k equal-sized sets. After that, the model is trained on k-1 sets and then tested on the remaining set. As a result, the model's accuracy varies. The final accuracy result is calculated by taking mean of the accuracy found for the folds. Our ABO\textsubscript{3} perovskite structure classification and perovskite prediction results are illustrated in the the Table \ref{table:1} and \ref{table:2} respectively.

\begin{table}[htb]
\centering
    \begin{tabular}{c|c|c|c|c}
      \textbf{Algorithm} & \textbf{Accuracy} & \textbf{Precision} & \textbf{Recall} & \textbf{F1}\\
      \hline
      Random Forest & 90.56\% & 91.29\% & 90.56\% & 90.2\%\\
      XGBoost & 89.72\% & 90.54\% & 89.72\% & 89.22\%\\
      LGBM & 87.15\% & 88.11\% & 87.15\% & 86.35\% \\
      Decision Tree & 88.9\% & 89.77\% & 88.97\% & 88.4\% \\
      KNN & 83.81\% & 83.73\% & 83.81\% & 83.05\%\\
    \end{tabular}
    \caption{Evaluation metrics for structure classifier model}
    \label{table:1}
\end{table}

\begin{table}[htb]
\centering
    \begin{tabular}{c|c|c|c|c}
      \textbf{Algorithm} & \textbf{Accuracy} & \textbf{Precision} & \textbf{Recall} & \textbf{F1}\\
      \hline
      Random Forest & 97.65\%\ & 97.65\% & 97.65\% & 97.65\%\\
      XGBoost & 96.34\% & 96.34\% & 96.34\% & 96.34\%\\
      LGBM & 96.86\% & 96.86\% & 96.86\% & 96.86\% \\
      Decision Tree & 97.12\% & 97.12\% & 97.12\% & 97.12\%\\
      KNN & 89.15\% & 89.15\% & 89.15\% & 89.15\%\\
    \end{tabular}
    \caption{Evaluation metrics for formability prediction model}
    \label{table:2}
\end{table}
Random Forest (RF) algorithm outperformed all the used algorithms. Feature selection and explanation are done for Random Forest in the subsequent section. 

\subsection{SHAP}
It is essential to know which features influenced our random forest model's classification choices and which features led to a particular prediction outcome. To determine the effects of the different features in our prediction, we used TreeExplainer's SHAP (SHapley Additive exPlanations) implementation~\cite{lundberg2020local}.
\begin{figure}[!htb]
\centering
    \includegraphics[width=0.7\linewidth]{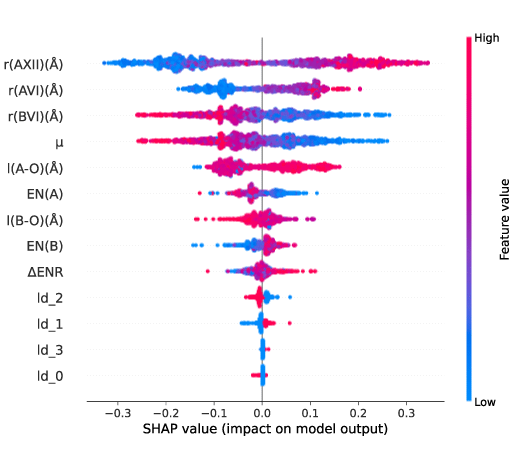}
    \caption{Feature vs. Shapley value plot for perovskite formability prediction}
    \label{fig:shap form}
\end{figure}
\begin{figure*}[!htb]
\centering
    \begin{subfigure}[hbt]{0.245\linewidth}
    \centering
    \includegraphics[width=\linewidth]{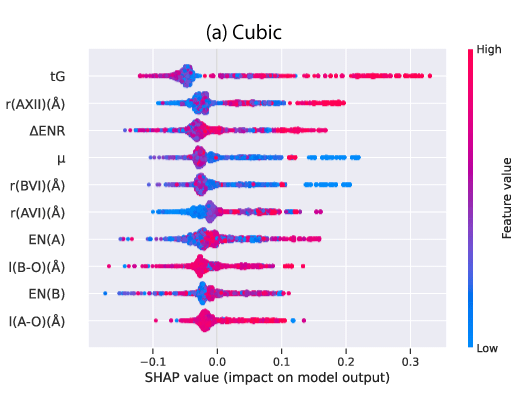}
    \end{subfigure}
    \hfill
    \begin{subfigure}[hbt]{0.245\linewidth}
    \centering
    \includegraphics[width=\linewidth]{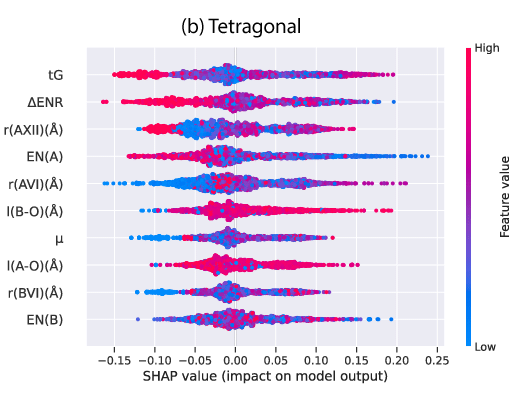}
    \end{subfigure}
    \hfill
    \begin{subfigure}[hbt]{0.245\linewidth}
    \centering
    \includegraphics[width=\linewidth]{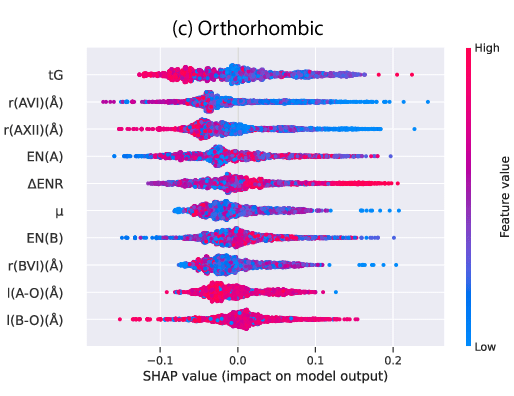}
    \end{subfigure}
    \hfill
    \begin{subfigure}[hbt]{0.245\linewidth}
    \centering
    \includegraphics[width=\linewidth]{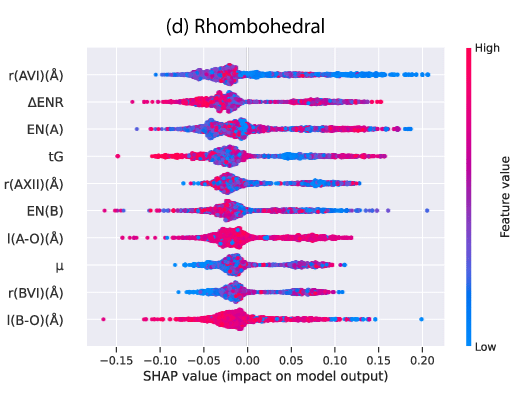}
    \end{subfigure}
\caption{Feature vs. Shapley value plot for perovskite structure classification}
\label{fig:shap class}
\end{figure*}
The SHAP summary plots (features vs. shapley values) are visualized in Fig.(\ref{fig:shap form},\ref{fig:shap class}). Positive shapley values on the horizontal axis suggest that the particular class is likely to be anticipated for that feature. The features on the vertical axis are listed in descending order based on their impact on the prediction. In the plots, the red hue denotes a high value for a feature at the data point, and blue indicates a low feature value. As the feature which is significantly contributing to a class can be pointed out using the shapley values, that particular feature may be manipulated based on necessity while developing the material. It is to be remembered that SHAP values aid in determining the contribution of features to the classification; nevertheless, they do not ensure causation between feature value and prediction probability.

From the SHAP summary plot of perovskite formability prediction, we gain insight that the most important feature is the radius of Cation A with 12 fold coordination. A high value of $r(AXII)$ has the most positive impact on the perovskite formability prediction model. Shannon radii for the A and B cations are chosen to be closest to 12 and 6, which are consistent with the A and B cations' coordination environments in the perovskite structure~\cite{bartel2019new}. Octahedral factor is another feature that plays important role in the formation of perovskite. Low value of the octahedral factor has a positive impact on the prediction model. This is also in line with the fact that a cubic perovskite structure requires an octahedral factor ($r(B)/r(O)$) value of 0.414~\cite{kumar2008prediction}. 

For perovskite crystal structure prediction from the summary plot, we can state that high value of $t_G$ and high value $r(AXII)$ has a positive impact on the model for predicting cubic structures. A $t_G$ value of unity suggests a completely cubic structure according to Goldschmidt's hard-sphere model. This assumption holds truer in structures with high ionicity in the bonding. High value of radius of cation A with 12 fold coordination has a high positive impact on the model. Also, high value of $t_G$ is good for classifying perovskites.
 
For predicting tetragonal structure, a high value of $l(B-O)$ and low value of $EN(A)$ has a positive impact on the model. If $l(B-O)$ values are between 1.957–2.174\si{\angstrom}, it helps to form $A^+B^{5+}O^{3-}$ type perovskite compounds. For $A^{2+}B^{4+}O^3$ type perovskite compounds the $l(B-O)$ value should be between 1.870–2.330\si{\angstrom}. This is seen in the case of CsNbO\textsubscript{3} that does not form a perovskite structure because of small $l(B-O)$ value~\cite{zhang2007structural}. The covalent interaction between B-O is greater than that of A-O in the ABO\textsubscript{3} structure, resulting in lattice deformation and thus the phase change (cubic- tetragonal). In tetragonal structures, a high EN(A) value is preferable since it indicates a more covalent character. Additionally, the high values of ${t_G}$ cause the B cation to rattle inside the tetragonal unit cell, resulting in its ferroelectric characteristic~\cite{behara_crystal_2021}.

For predicting orthorhombic structures, the high value of $\Delta{ENR}$ and low value of $r(AXII)$ has a positive impact on the model. In the case of predicting rhombohedral structures, the high value of $t_G$ and low value of $r(AVI)$ has a positive impact on the model.

\section{Conclusion}
The paper primarily focused on classifying the crystal structure of the ABO\textsubscript{3} perovskite and predicting the formability of perovskites. Using our parameter tuned RF model, we have achieved 90.56\% accuracy in classifying perovskite crystal structures and 97.65\% accuracy in predicting perovskite crystal formability. Which performs significantly well than the previous studies. Additionally, Shapley analysis is used to figure out and select the importance of features for the models. As the impact of different features is not conclusive yet, this analysis may provide additional insight on the impact of different features. Our work will make it easier to predict the formability and structure of perovskites with relatively constrained features using our model. The crystal structure classification is important as the stability of perovskite heavily depends on it. For future work, we intend to concentrate on the bandgap of the perovskite material, as different bandgaps are used for different applications. Our world is shifting towards renewable energy sources, discovery of better perovskite material will greatly accelerate the efficiency and commercially implementable of PV panels. Further research is required to gain a better understanding of perovskites' formability and structure.

\printbibliography

\end{document}